\documentclass[prl,twocolumn,showpacs,aps,superscriptaddress,floatfix]{revtex4}
\usepackage{amsmath}
\usepackage{amssymb}
\usepackage{amsfonts}
\usepackage{bm}
\usepackage{graphicx}
\usepackage{color}
\usepackage{hyperref}
\usepackage{verbatim}

\begin{document}

\title{Superconductivity via repulsive interaction in the model with electrons and holes}

\author{A.O. Sboychakov}
\affiliation{Institute for Theoretical and Applied Electrodynamics, Russian Academy of Sciences, Moscow, 125412 Russia}

\begin{abstract}
We study a toy model of the 2D electron gas including electron-like and hole-like charge carriers hybridized between each other. Electrons and holes can move either in different layers or in the same layer. Undoped system is an insulator. Slight doping brings extra charge carriers having large density of states. We show that the screened Coulomb interaction is enough to stabilize the superconductivity in the slightly doped system. The symmetry of the order parameter corresponds to the spin-triplet $p$-wave order. For realistic values of the model parameters the superconducting transition temperature can reach the value of $150$ kelvin.
\end{abstract}

\date{\today}

\maketitle

\section{Introduction}

As it was shown by W.~Kohn and J.M.~Luttinger in Ref~\cite{KohnLuttinger1965} the screening of the Coulomb interaction in fermion systems leads to the superconductivity. This means that any Fermi-liquid not only with attraction but also with repulsion is unstable with respect to the superconductivity. At the same time, the estimates of the transition temperature, $T_c$, showed that for the majority of metals $T_c$ is too small and often lays far below experimentally achievable values (see, however, the paper~\cite{KaganChubukovHe3} showing that the transition temperature for $^3$He calculated in the framework of the Kohn--luttinger mechanism is consistent with experimental data). However, the recent theoretical studies of the graphene multilayers demonstrate~\cite{KLSC2019,wagner2023superconductivity,OurTripletPRB2023,KLSCtrilayerPRB2022Guinea,RPAtetralayerPRL2025,SCMeleV2026} that the transition temperature calculated in the framework of the Kohn--Luttinger mechanism turn out to be large enough to be experimentally detected. Moreover, the transition temperature calculated for the AB bilayer graphene~\cite{OurTripletPRB2023}, rhombohedral trilayer~\cite{KLSCtrilayerPRB2022Guinea}, and tetralayer~\cite{RPAtetralayerPRL2025} graphenes turned out to be in quantitative agreement with experimentally observable values (see also review paper~\cite{kagan2025}).

Motivated by these studies we propose here a simple two-dimensional (2D) toy model, which exhibits the superconducting instability with large enough transition temperature. The model includes electron-like and hole-like charge carriers with hybridization between them. The undoped 2D system is a band insulator. Small doping brings extra charge carriers into the system with large density of states. We show that the screened Coulomb interaction, calculated within random phase approximation (RPA) stabilizes the superconducting state in this system. The symmetry of the order parameter is of the spin-triplet $p$-wave type. For realistic values of the model parameter the superconducting transition temperature can be about $150$ kelvin.

\section{Single-particle model Hamiltonian}

Let us consider some two-dimensional system consisting of one or several layers and let us assume that the single-particle Hamiltonian of this system has a form
\begin{eqnarray}\label{H0}
H_0&=&\sum_{\mathbf{p}f\sigma}\left(\varepsilon_{e\mathbf{p}}d^{\dag}_{\mathbf{p}ef\sigma}d^{\phantom{\dag}}_{\mathbf{p}ef\sigma}+
\varepsilon_{h\mathbf{p}}d^{\dag}_{\mathbf{p}hf\sigma}d^{\phantom{\dag}}_{\mathbf{p}hf\sigma}\right)+\nonumber\\
&+&\sum_{\mathbf{p}f\sigma}\left(\Delta_0d^{\dag}_{\mathbf{p}ef\sigma}d^{\phantom{\dag}}_{\mathbf{p}hf\sigma}+H.c.\right)\,,
\end{eqnarray}
where $d^{\dag}_{\mathbf{p}\alpha f\sigma}$ ($d^{\phantom{\dag}}_{\mathbf{p}\alpha f\sigma}$) is the creation (annihilation) operator of the electron with momentum $\mathbf{p}$, "electron-hole" index $\alpha$($=e,h$), flower (e.g., valley) index $f$($=1,2,\dots,N_f$), and spin projection $\sigma$. If charge carriers with different $\alpha$ move in different layers (without hybridization) then index $\alpha$ corresponds also to the layer index. However, we can also suppose that all the charge carriers live in the same layer. In the latter case, the index $\alpha$ just distinguishes the electron-like and hole-like charge carriers. We will consider both these possibilities in the same manner. The function $\varepsilon_{\alpha\mathbf{p}}$ describes the dispersion of the charge carriers. Below we assume for simplicity that
\begin{equation}\label{varepsilon12}
\varepsilon_{e\mathbf{p}}=-\varepsilon_{h\mathbf{p}}\equiv\varepsilon_{\mathbf{p}}=\frac{\hbar^2p^2}{2m^{*}}-\frac{\hbar^2p_0^2}{2m^{*}}\,,
\end{equation}
where $m^{*}$ is the effective mass and $p_0$ is the Fermi momentum for uncoupled ($\Delta_0=0$) charge carriers. Below we assume that $m^{*}\sim m_e$ and $p_0\sim1/a$, where $m_e$ is the electron mass and $a$ is the intra-layer lattice constant. The function $\varepsilon_{\mathbf{p}}$ can also be rewritten in the form
\begin{equation}\label{varespilon}
\varepsilon_{\mathbf{p}}=E^{*}\left[(pa)^2-(p_0a)^2\right]\,,\;\;\;E^{*}=\frac{\hbar^2}{2m^{*}a^2}\,.
\end{equation}
We also introduce the energy $E_0=\hbar^2/(2m_ea_0^2)$ as a reference energy, where we take $a_0=1.5$\,\AA. With this value, we have $E_0\cong1.7$\,eV. Below we assume also that $a\sim a_0$, so the $E_0$ is the characteristic energy for $\varepsilon_{\mathbf{p}}$.

\begin{figure}[t]
\includegraphics[width=0.99\columnwidth]{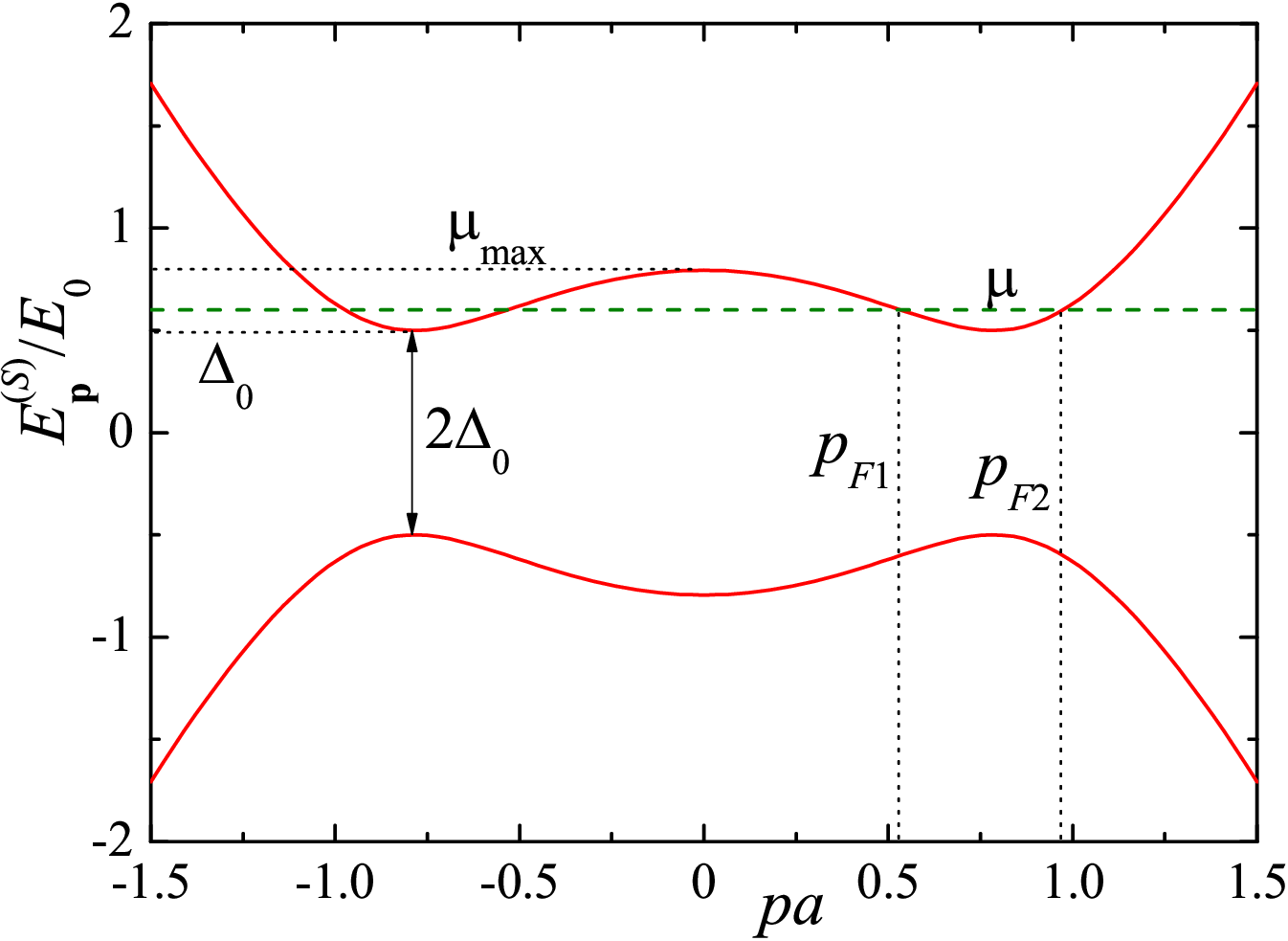}
\caption{\label{FigSpec} The electron spectrum, Eq.~\eqref{Spec}, calculated at $E^{*}=E_0$, $\Delta_0=0.5E_0$, $a=a_0$, and $p_0=\pi/(4a_0)$. In doped system the chemical potential, laying in the range $\Delta_0<\mu<\sqrt{\Delta_0^2+p_0^4a^4E^{*2}}\equiv\mu_{\text{max}}$, forms two circular Fermi surface sheets, with Fermi momenta $p_{F1,2}$ given by Eq.~\eqref{pF12}.}
\end{figure}

The second term in Eq.~\eqref{H0} describes the hybridization of electrons and holes. The energy spectrum of the Hamiltonian~\eqref{H0} is
\begin{equation}\label{Spec}
E^{(1,2)}_{\mathbf{p}}=\mp\sqrt{\Delta_0^2+\varepsilon_{\mathbf{p}}^2}\,.
\end{equation}
The typical curves $E^{(1,2)}_{\mathbf{p}}$ vs $p$ are shown in Fig.~\ref{FigSpec}. The spectrum has a gap, so the undoped system is the band insulator with the gap $2\Delta_0$. Below we assume that $\Delta_0\sim E_0$. The eigenfunctions corresponding to the bands~\eqref{Spec} are:
\begin{equation}\label{Phi}
\Phi^{(1,2)}_{\mathbf{p}\alpha}=\frac{1}{\sqrt{2}[\varepsilon_{\mathbf{p}}^2+\Delta_0^2]^{1/4}}\left(
\begin{array}{c}
\mp\sqrt{\sqrt{\varepsilon_{\mathbf{p}}^2+\Delta_0^2}\mp\varepsilon_{\mathbf{p}}}\\
\displaystyle\frac{\Delta_0}{\sqrt{\sqrt{\varepsilon_{\mathbf{p}}^2+\Delta_0^2}\mp\varepsilon_{\mathbf{p}}}}
\end{array}
\right)\,.
\end{equation}

For the reason discussed below, we also assume that the electron operator $d^{\phantom{\dag}}_{\mathbf{p}\alpha f\sigma}$ posses the flower index $f$, ranging from $1$ to $N_f$. So, the model includes $N_f$ copies of the spectra~\eqref{Spec}. We also introduce the band electronic operators defined as $\gamma_{\mathbf{p}Sf\sigma}=\sum_{\alpha}\Phi^{(S)}_{\mathbf{p}\alpha}d_{\mathbf{p}\alpha f\sigma}$. In terms of these operators the Hamiltonian~\eqref{H0} becomes diagonal one.

In this paper we consider slightly doped system, when the chemical potential $\mu$ lays in the range $\Delta_0<\mu<\sqrt{\Delta_0^2+p_0^4a^4E^{*2}}$. With this restriction, the Fermi surface consists of the two circular sheets with the Fermi momenta
\begin{equation}\label{pF12}
p_{F1,2}=\frac{1}{a}\sqrt{(p_0a)^2\mp\sqrt{\mu^2-\Delta_0^2}/E^{*}}\,.
\end{equation}
The density of states at the Fermi level is
\begin{equation}\label{DOS}
\rho(\mu)=\frac{N_f}{\pi a^2}\frac{\mu}{E^{*}\sqrt{\mu^2-\Delta_0^2}}\,.
\end{equation}
The density of states diverges when the chemical potential goes to the band edge. As we will show below this stabilizes the superconductivity mediated by the screened Coulomb interaction.

\section{Polarization operator and screened Coulomb interaction}

Now we add the interaction into our model. We choose the interaction Hamiltonian in the form
\begin{equation}\label{Hint}
H_{\text{int}}=\frac{1}{2\cal N}\!\!\sum_{\mathbf{pp}'\mathbf{q}}\!\sum_{\alpha\beta ff'\atop\sigma\sigma'}\!
d^{\dag}_{\mathbf{p}+\mathbf{q}\alpha f\sigma}d^{\phantom{\dag}}_{\mathbf{p}\alpha f\sigma}
V_{0\mathbf{q}}d^{\dag}_{\mathbf{p}'-\mathbf{q}\beta f'\sigma'}d^{\phantom{\dag}}_{\mathbf{p}'\beta f'\sigma'}\,.
\end{equation}
The function $V_{0\mathbf{q}}$ here is the "bare" interaction function. In the framework of our toy model we choose this function in the simple form
\begin{equation}\label{V0}
V_{0\mathbf{q}}=\frac{2\pi e^2}{\epsilon q},
\end{equation}
which is just the 2D Fourier transform of the Coulomb interaction. In this equation, $\epsilon$ is the dielectric constant of the media surrounded the sample. Let us notice that in real materials the function $V_{0\mathbf{q}}$ can deviate from Eq.~\eqref{V0}. Thus, it can include the wave functions which diagonalize the first term of the Hamiltonian~\eqref{H0}. It also may depend on the "electron-hole" indices. We introduce two dimensionless quantities: $\alpha^{*}=e^2/(\epsilon E^{*}a)$ and $\alpha_0=e^2/(E_0a_0)$. For $a_0$ chosen, we have $\alpha_0\cong5.66$. Since we assume here that $E^{*}\sim E_0$ and $a\sim a_0$, we have $\alpha^{*}\sim\alpha_0$ if the dielectric constant $\epsilon$ is not very large. Thus, for the energy parameters used we are on the strong coupling regime, when $\alpha^{*}>1$.

\begin{figure}[t]
\includegraphics[width=0.99\columnwidth]{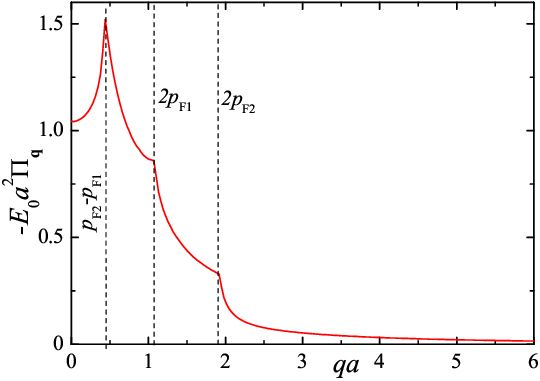}
\caption{\label{FigP} Momentum dependence of the polarization operator calculated at $\mu/\Delta_0=1.05$. Other model parameters are: $E^{*}=E_0$, $\Delta_0=E_0$, $a=a_0$, and $p_0=\pi/(4a_0)$.}
\end{figure}

\begin{figure}[t]
\includegraphics[width=0.99\columnwidth]{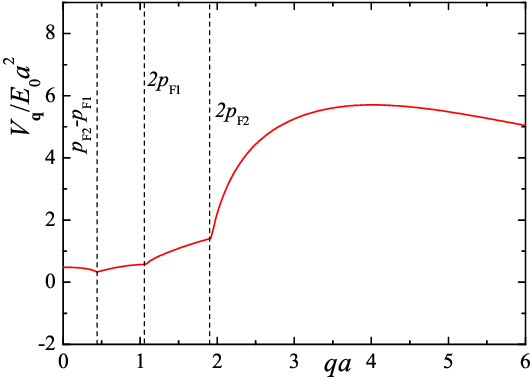}
\caption{\label{FigV} Momentum dependence of the screened Coulomb interaction calculated at $\alpha=\alpha_0$. Other model parameters are the same as in Fig.~\ref{FigP}.}
\end{figure}

Since the doped system is a metal with large density of states, the Coulomb interaction at small transfer momenta experiences large screening. We cannot use here the perturbation theory to calculate the screening since we have $\alpha^{*}>1$. For this reason we calculate the screened Coulomb interaction within RPA. This is uncontrollable approximation. However, in all calculations below we take $N_f=2$, as in graphene-based materials. Thus, each bubble diagram enters with weight $2N_f=4$, which, as as expected~\cite{InteractionsGrapheneReview2012}, makes RPA an appropriate approach.

Within RPA the screened Coulomb interaction is
\begin{equation}\label{VRPA}
V_{\mathbf{q}}=\frac{V_{0\mathbf{q}}}{1-\Pi_{\mathbf{q}}V_{0\mathbf{q}}},
\end{equation}
where $\Pi_{\mathbf{q}}$ is the static polarization operator of the system. It can be written as
\begin{eqnarray}\label{P}
\Pi_\mathbf{q}&=&2N_f\sum_{SS'}\int\!\frac{d^2\mathbf{p}}{(2\pi)^2}
\frac{n_F\left(E^{(S)}_{\mathbf{p}}\right)-n_F\left(E^{(S')}_{\mathbf{p}+\mathbf{q}}\right)}
{E^{(S)}_{\mathbf{p}}-E^{(S')}_{\mathbf{p}+\mathbf{q}}}\times\nonumber\\
&\times&\Big(\sum_{\alpha}\Phi^{(S)}_{\mathbf{p}\alpha}\Phi^{(S')}_{\mathbf{p}+\mathbf{q}\alpha}\Big)
\Big(\sum_{\beta}\Phi^{(S)}_{\mathbf{p}\beta}\Phi^{(S')}_{\mathbf{p}+\mathbf{q}\beta}\Big),
\end{eqnarray}
where $n_F(E)$ is the Fermi function.

We calculated $\Pi_\mathbf{q}$ numerically at $T=0$ for different values of the single-particle model parameters and different values of the chemical potential. Typical curve is shown in Fig.~\ref{FigP}. At $0<q<p_{F2}-p_{F1}$, the $-\Pi_{\mathbf{q}}$ increases with $q$, while in the region $p_{F2}-p_{F1}<q\lesssim2p_{F2}$ it decreases fast with $q$. We clearly see three Kohn anomalies located at $q=p_{F2}-p_{F1}$, $q=2p_{F1}$, and $q=2p_{F2}$. At $q>2p_{F2}$, the polarization operator behaves as for usual 2D electron gas with quadratic dispersion.

The momentum dependence of the screened Coulomb interaction is shown in Fig.~\ref{FigV}. The $V_{\mathbf{q}}$ increases with $q$ at $q\lesssim2p_{F2}$. Such a behavior will stabilize the superconducting order.

\section{Superconductivity}

The interaction Hamiltonian in Cooper channel and in terms of the band electronic operators can be written as
\begin{equation}\label{HintBCS}
H_{\text{int}}=\frac{1}{2\cal N}\!\!\sum_{\mathbf{pp}'}\!\sum_{f\sigma\sigma'}
\gamma^{\dag}_{\mathbf{p}2f\sigma}\gamma^{\dag}_{-\mathbf{p}2f\sigma'}
\Gamma_{\mathbf{pp}'}\gamma^{\phantom{\dag}}_{-\mathbf{p}'2f\sigma'}\gamma^{\phantom{\dag}}_{\mathbf{p}'2f\sigma}\,,
\end{equation}
where
\begin{equation}\label{Gamma}
\Gamma_{\mathbf{pp}'}=\Big(\sum_{\alpha}\Phi^{(2)}_{\mathbf{p}\alpha}\Phi^{(2)}_{\mathbf{p}'\alpha}\Big)V_{\mathbf{p}-\mathbf{p}'}
\Big(\sum_{\beta}\Phi^{(2)}_{-\mathbf{p}\beta}\Phi^{(2)}_{-\mathbf{p}'\beta}\Big)\,.
\end{equation}
In Eq.~\eqref{HintBCS} we kept only the terms, which open the gap at the Fermi level. We also assume that no inter-flower coupling arise, so the expectation values $\langle\gamma^{\phantom{\dag}}_{-\mathbf{p}2f\sigma}\gamma^{\phantom{\dag}}_{\mathbf{p}2f'\sigma'}\rangle=0$ if $f\neq f'$. However, for $f=f'$ some of the expectation values of the type $\langle\gamma^{\phantom{\dag}}_{-\mathbf{p}2f\sigma}\gamma^{\phantom{\dag}}_{\mathbf{p}2f\sigma'}\rangle$ are non-zero. We introduce the order parameter
\begin{equation}\label{DeltaDef}
\Delta_{\mathbf{p}\sigma\sigma'}=\sum_{\mathbf{p}'}\Gamma_{\mathbf{pp}'}
\langle\gamma^{\phantom{\dag}}_{-\mathbf{p}'2f\sigma}\gamma^{\phantom{\dag}}_{\mathbf{p}'2f\sigma'}\rangle\,,
\end{equation}
where we assume that $\Delta_{\mathbf{p}\sigma\sigma'}$ does not depend on the flower index. The spin structure of the order parameter is different for the spin-singlet and spin-triplet superconducting states. For the spin-singlet state, we have $\Delta_{\mathbf{p}\sigma\sigma'}=\Delta_{\mathbf{p}}[i\sigma^{y}_{\sigma\sigma'}]$, while for the spin-triplet state we have $\Delta_{\mathbf{p}\sigma\sigma'}=\bm{\Delta}_{\mathbf{p}}[i\sigma^{y}\bm{\sigma}]_{\sigma\sigma'}$. Below we assume the following anzac for the vector-valued spin-triplet order parameter: $\bm{\Delta}_{\mathbf{p}}=\Delta_{\mathbf{p}}\mathbf{n}$, where $\mathbf{n}$ is arbitrary unit vector. In this case, both spin-singlet and spin-triplet states can be considered in similar manner. However, the symmetrical properties of the function $\Delta_{\mathbf{p}}$ are different for spin-singlet and spin-triplet states. Thus, we have
\begin{equation}\label{DeltaSym}
\Delta_{-\mathbf{p}}=\pm\Delta_{\mathbf{p}}\,,
\end{equation}
where plus (minus) sign corresponds to the spin-singlet (spin-triplet) state.

Using the standard technique we obtain the self-consistency equation for the order parameter:
\begin{equation}\label{DeltaEq}
\Delta_{\mathbf{p}}\!=-\!\int\!\frac{d^2\mathbf{p}'}{(2\pi)^2}\Gamma_{\mathbf{pp}'}\Delta_{\mathbf{p}'}
\frac{\tanh\left[\frac{\sqrt{(E_{\mathbf{p}'}-\mu)^2+\Delta_{\mathbf{p}'}^2}}{2T}\right]}{2\sqrt{(E_{\mathbf{p}'}-\mu)^2+\Delta_{\mathbf{p}'}^2}}\,,
\end{equation}
where $E_{\mathbf{p}}=E^{(2)}_{\mathbf{p}}$. At $T\to T_c$, this equation can be linearized. As a result, we obtain
\begin{equation}\label{DeltaEqTtoTc}
\Delta_{\mathbf{p}}=-\!\int\!\frac{d^2\mathbf{p}'}{(2\pi)^2}\,\Gamma_{\mathbf{pp}'}\Delta_{\mathbf{p}'}
\frac{\tanh\left[(E_{\mathbf{p}'}-\mu)/(2T)\right]}{2(E_{\mathbf{p}'}-\mu)}\,.
\end{equation}
From this equation we find both the $T_c$ and the momentum dependence of the order parameter at $T\to T_c$ using the technique described in Ref.~\cite{SCMeleV2026}. Namely, we observe that the following relation holds true
\begin{equation}\label{GammaSym}
\Gamma_{\mathbf{pp}'}=\Gamma(p,p',\cos(\varphi_{\mathbf{p}}-\varphi_{\mathbf{p}'}))\,,
\end{equation}
where $\varphi_{\mathbf{p}}$ is the polar angle of the vector $\mathbf{p}$. Thus, we can seek a solution to Eq.~\eqref{DeltaEqTtoTc} in the form $\Delta_{\mathbf{p}}=\Delta_{p}\cos(\ell\varphi_{\mathbf{p}})$, where $\ell=1,2\,\dots$. Further, we introduce the quantity $\delta_p=\Delta_pf_p$ with
\begin{equation}\label{f}
f_p=\sqrt{p\frac{\tanh\left[(E_{p}-\mu)/(2T)\right]}{2(E_p-\mu)}}\,.
\end{equation}
The equation for $\delta_p$ is
\begin{equation}\label{deltaEq}
\delta_p=\int\limits_{0}^{p_{\text{max}}}\!dp\,w^{(\ell)}(p,p')\delta_{p'}
\end{equation}
with
\begin{equation}\label{wl}
w^{(\ell)}(p,p')=-\frac{f_pf_{p'}}{(2\pi)^2}\int\limits_{0}^{2\pi}d\varphi\,\Gamma(p,p',\cos\varphi)\cos(\ell\varphi)\,.
\end{equation}
The kernel $w^{(\ell)}(p,p')$ is symmetric one, so its eigenvalues are real. The transition temperature for given $\ell$ is found as the point where the largest eigenvalue of $w^{(\ell)}(p,p')$ is equal to $1$. In our numerical scheme we use linear mesh with $N_p=800$ momentum points in the region $0<p<p_{\text{max}}=\pi/a$. We obtain that for all model parameters used the leading instability corresponds to the $p$-wave ($\ell=1$), that is, the order parameter has the form
\begin{equation}\label{DeltaPwave}
\Delta_{\mathbf{p}}=\Delta_{p}\cos\varphi_{\mathbf{p}}\,.
\end{equation}
Obviously, this order parameter corresponds to the spin-triplet state.

\begin{figure}[t]
\includegraphics[width=0.99\columnwidth]{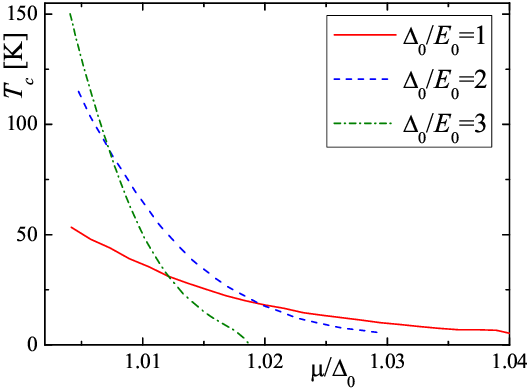}
\caption{\label{FigTcvsMu} The dependence of the transition temperature on the chemical potential calculated for three different values of $\Delta_0$. Model parameters are: $E^{*}=E_0$, $a=a_0$, $p_0=\pi/(4a_0)$, $\alpha=\alpha_0$.}
\end{figure}

We calculated $T_c$ for different model parameters. Figure~\ref{FigTcvsMu} shows the results for the $\mu$ dependence of the transition temperature calculated for three different values of $\Delta_0$. Other model parameters are: $E^{*}=E_0$, $a=a_0$, $p_0=\pi/(4a_0)$, $\alpha=\alpha_0$. We see that at small $\mu$, the larger $\Delta_0$ is, the large is $T_c$. However, for large $\mu$ the situation is opposite. For $\Delta_0=3E_0\cong5.2$\,eV, the transition temperature reaches the value of $150$\,K. The transition temperature increases with the decrease of $\mu$. This is because the density of states~\eqref{DOS} goes to infinity when $\mu\to\Delta_0$. In Fig.~\ref{FigTcvsMu} we show only the results for $T_c(\mu)>\varepsilon_F(\mu)=\mu-\Delta_0$. At smaller $\mu$ we should take into account the temperature dependence of the polarization operator, which substantially increases the computation costs. However, it is clear that at small $\mu$ the superconductivity will be suppressed by the impurities.

\begin{figure}[t]
\includegraphics[width=0.99\columnwidth]{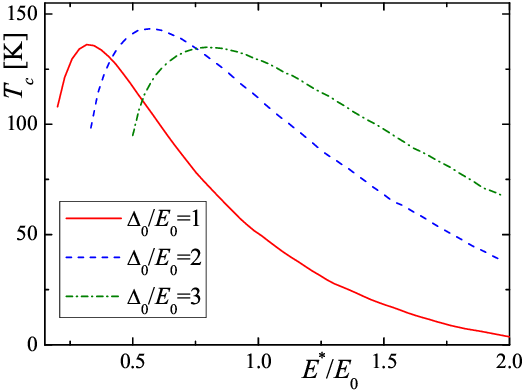}
\caption{\label{FigTcvsEs} The dependence of the transition temperature on the parameter $E^{*}$ calculated for three different values of $\Delta_0$. Calculations are performed at fixed ratio $\varepsilon(\mu)/\Delta_0=\mu/\Delta_0-1=0.005$. Other model parameters are: $a=a_0$, $p_0=\pi/(4a_0)$, $\epsilon=1$.}
\end{figure}

Figure~\ref{FigTcvsEs} shows the dependence of the transition temperature on the parameter $E^{*}$ calculated at three different values of $\Delta_0$. Calculation are performed at fixed ratio $\mu/\Delta_0=1.005$. The $T_c$ increases with $E^{*}$ at small $E^{*}$ and decreases with $E^{*}$ at large $E^{*}$. Such a behavior of $T_c$ seems quite natural. Indeed, the transition temperature can be presented in the form $T_c\propto E^{*}e^{-1/\lambda(E^{*})}$, where $\lambda(E^{*})$ is the dimensionless coupling constant. When $E^{*}\to0$, the coupling constant $\lambda$ goes to some finite value, and the transition temperature will decrease with the decrease of $E^{*}$ due to the preexponential factor. At large $E^{*}$, the coupling constant decreases with the increase of $E^{*}$ since the parameter $\alpha^{*}=e^2/(\epsilon E^{*})$ decreases with  $E^{*}$.

For all tree curves shown in Fig.~\ref{FigTcvsEs} the maximum value of $T_c$ is slightly below $150$\,K. One may assume that the value about $150$\,K is the maximum value of $T_c$ achievable in our model. Indeed, for the results presented in Fig.~\ref{FigTcvsEs} we have $\max(T_c)\approx\varepsilon_F(\mu)$, that is, we are at the edge of the applicability of zero temperature polarization operator approximation. Further decrease of $\mu$ requires the calculation of the polarization operator at finite temperatures. This region is outside of our numerical procedure. However, one can expect, that at $\mu\to\Delta_0$ the transition temperature will go to zero since the undoped system is an unsulator.

\begin{figure}[t]
\includegraphics[width=0.99\columnwidth]{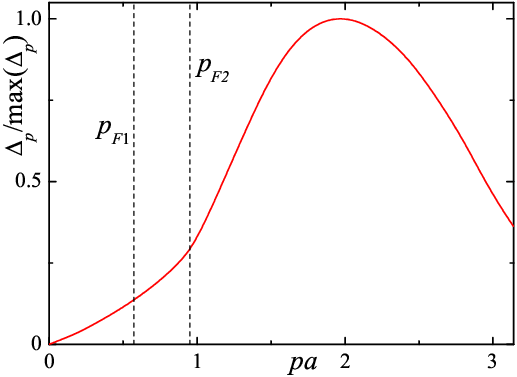}
\caption{\label{FigDelta} The momentum dependence of the order parameter $\Delta_{p}$. Model parameters are: $\Delta_0=2E_0$, $E^{*}=E_0$, $a=a_0$, $p_0=\pi/(4a_0)$, $\mu/\Delta_0=1.01$, $\alpha=\alpha_0$.}
\end{figure}

The approach used allows also to calculate the momentum dependence of the order parameter at $T\to T_c$. Namely $\Delta_p=\delta_p/f_p$, where $\delta_p$ is the eigenfunction of the kernel $w^{(1)}(p,p')$ corresponding to its largest eigenvalue. Typical curve is shown in Fig.~\ref{FigDelta}. The function $\Delta_p$ has a maximum at $p\approx2p_{F2}$, while at $p=p_{F1,2}$ it is several times smaller then its maximum value.

\section{Discussion}

Thus, in the framework of our toy model including electron-like and hole-like charge carries we showed that the screened Coulomb interaction stabilizes the spin-triplet $p$-wave superconducting order. The critical temperature can reach the value about $150$\,K. Let us proposed here the hypothetical 2D system for which the model used can be applicable (with some modifications). Namely, let us consider three layer system with square lattice in each layer. The top and bottom layers are metallic ones. Taking into account only the nearest-neighbor intra-layer hopping (with amplitude $t$), the dispersion of the charge carriers in these two layers is $\varepsilon_{\mathbf{p}}=-2t[\cos(p_xa)+\cos(p_ya)]$. Let us also assume that the intermediate layer is dielectric one with chess-like charge density wave order, which (i) doubles the lattice period and (ii) modulates the electron hopping between the top and bottom layers. The superlattice of the system contains two square sublattices $A$ and $B$. Let us suppose that the amplitude of the electron hopping between nearest-neighbor sites of the top and bottom layers in sublattice $A$ is equal to $t_A=\Delta_0$, while the same quantity in the sublattice $B$ is equal to $t_B=-\Delta_0$. One can show that in this case the electron spectrum of the system will be two-fold degenerate ($N_f=2$), and it will be described by Eq.~\eqref{Spec}. Thus, the only difference of the electron spectrum from our toy model will be in the form of the dispersion law of $\varepsilon_{\mathbf{p}}$. Note also that for the proposed hypothetical system the singularity of the density of states at $\mu\to\Delta_0$ will be even stronger then for the toy model considered here, because at $E=0$ the proposed three layer system posses the van Hove singularity (if only nearest-neighbor hopping is taken into account). The way of how to reach the condition $t_A=-t_B$ is unknown for us. Note, however, that, perhaps, this condition can be weakened (from view point of the value of $T_c$).

\section{Conclusions}

In conclusion, we proposed the 2D toy model with hybridized electron-like and hole-like charge carriers, which can move in the same or different layers. At no doping the system is unsulator. Slight doping brings extra charge carriers with large density of states, which diverges when the chemical potential goes to the band edge. We showed that the screened Coulomb interaction stabilizes the spin-triplet $p$-wave superconducting order. We analyzed the dependence of $T_c$ on the model parameters. We showed the for the realistic values of model parameter the critical temperature can reach the value of about $150$\,K. 

The author is grateful A.L.~Rakhmanov and A.V.~Rozhkov for useful discussions.



\end{document}